\documentclass[12pt]{article}

\begin{document}

\title{
{\it AquaFuel},
An  Example  Of  The  Emerging  New  Energies
And  The  New  Methods  For  Their
Scientific  Study }
\author{Ruggero Maria Santilli}
\date{ }
\maketitle
\begin{center}
Institute for Basic Research \\
P. O. Box 1577, Palm Harbor, FL 34682, U.S.A. \\
ibr@gte.net, http://home1.gte.net/ibr/ \\
Received April 15, 1998 \\
\end{center}

\begin{abstract}

In this paper we  initiate systematic studies on the novel methods
needed for quantitative scientific studies of the emerging new forms of
energy, by using as a representative example  the new combustible gas
called AquaFuel, discovered and patented by William H. Richardson, jr.,
whose rights are now owned by Toups Technology Licensing, Inc. (TTL), of
Largo, Florida. In essence, AquaFuel is a new energy converter capable
of transforming Carbon and water into a new combustible gas via the use
of a suitable electric discharge. We show that AquaFuel can be produced
easily, safely and rapidly in large amounts, has novel physical and
chemical characteristics, and exhibits greatly reduced emission
pollutants as compared to fuels currently used. We then review nine
basic experimental measures currently under study by TTL which are
necessary for a scientific appraisal of AquaFuel. We outline the
limitations of quantum mechanics and chemistry for the treatment of
{\it new} forms of energies, namely, energies which by definition should
be
{\it beyond} said theories. We point out the availability of broader
theories specifically constructed for the study of new energies and
point out available applications. Detailed studies on the origin of the
AquaFuel energy and its energy balance  will be presented at a later
time upon completion of the systematic experimental and theoretical
studies currently under way. These results are expected to be
significant for various other forms of new energies.

\end{abstract}

\section{ The  Basic  Process  Underlying  AquaFuel}

AquaFuel is a new combustible gas discovered and developed by William H.
Richardson, jr, which is covered by the patents listed in Refs. [1]. Its
rights are now owned by Toups Technology Licensing, Inc.  (TTL), a U. S.
corporation in Largo, Florida, which is continuing its development.
Starting from the three basic elements of nature, Hydrogen, Oxygen and
Carbon, AquaFuel is produced during an electric discharge across an arc
between carbon electrodes immersed in distilled, fresh or salt water.

        The patented process is basically different from electrolysis.
In fact,
in electrolysis, water must be complemented with electrolyte to carry
the charges from the negative to the positive poles of the power source.
In AquaFuel charges are instead stimulated by a sufficient voltage
difference. In the latter case, no addition of electrolyte chemicals is
needed to create a conducting path. As we shall see, this property of
AquaFuel is important to minimize pollutants in the various
applications.

        According to currently available scientific knowledge,
the main process
underlying AquaFuel is constituted by Carbon atoms breaking loose under
discharge from the carbon electrodes and forming particular bonds with
the water constituents, Oxygen and Hydrogen. The resulting new molecules
cool as they bubble up to the water's surface where they are collected
and stored for various usages. The resulting combustible gas is called
AquaFuel.

\section{ Description of Equipment}

As in Figure 1, a one-gallon fishbowl is three-quarters filled with
distilled, tap or salt water. Copper tubing conducts 24 volts continuous
current (DC) to the tips of two 0.25
diameter Carbon rods composed of
99\% pure graphite, which are immersed in said water, one of whose tips
is extended into a large diameter Carbon block.

        The electric arc tunnels thru the water from the tip of
the Carbon rod
to the large Carbon block. The AquaFuel gas cools while bubbling to the
water's surface where it is collected with an inverted funnel.

        The electric arc produces a local temperature of
the order of 5,000 $ ^{\circ}F$.
The process dissociates the water molecules by forming a local high
temperature plasma composed of generally ionized atoms of Carbon,

For a picture of AquaFuel (photo by Rob
Jaeger and Ken Lindfors) visit the web site
http://www.toupstech.com/html/toups\_-\_aqua\_fuel.html. Note
the thin Carbon rod and the large Carbon block and the bubbling
of the gas thru the water and its clean burning at the top.

Hydrogen and Oxygen, which are subjected to a number of physical and
chemical processes. The final product is given by the recombination of
said elements into a number of new molecules that cool down in the water
surrounding the discharge as they bubble to the collection equipment.

        AquaFuel can be most efficiently produced via the use of
continuous
current (DC) in the arc. The use of alternating current (AC) in the arc
is also possible, although with a reduced production rate.

        The AquaFuel equipment is an energy converter because it
converts the
energy originally available (electricity, Carbon and water) into
different  forms of energy.

        In addition, the AquaFuel equipment is a new energy converter
because
its underlying processes are basically novel, as shown later on.

        This paper deals only with the excellent characteristics of
AquaFuel as
a combustible gas without any consideration at this time of the ratio
between the input and output energy which will be considered in a future
paper upon completion of the needed experimental measures.

\section{ Description of AquaFuel}

The actual structure of AquaFuel as well as its main physical and
chemical characteristics are still essentially unknown at this writing
on scientific grounds because the needed experimental measures are still
in progress (see Sect. 5).

        The various studies completed until now have identified the
following
main characteristics:

        1) AquaFuel is lighter than air, because it continues to rise in
atmosphere; 

        2)  AquaFuel does not self--combust because of its low content
of Oxygen
(estimated at about 1\% );

        3) AquaFuel is expected to be composed of H2, CO, Oxygen and a
number
of hydrocarbons;

        4) AquaFuel can run existing internal combustion (IC) engines
with
minor modifications in the carburetor;

        5) AquaFuel has an astonishingly low pollutant content in its
exhaust
when compared to other fuels, such as gasoline, methane, coal, etc. ;

        6) No pollution control equipment is needed for burning
AquaFuel;

        7) Engine oils remain much cleaner when burning AquaFuel because
of
less pollutants discharged during operation, thus lasting longer and
reducing engine oil disposal;

        8) The main gas produced in burning AquaFuel is carbon dioxide
as it is
the case of methane, whose precipitation into a solid form is under
study;

        9) AquaFuel is a stable, permanent gas, not like the mixture of
Hydrogen and Oxygen emitted from ordinary electrolysis;

        10) AquaFuel can be stored in ordinary tanks either as a gas or
in its
liquid form;

        11) AquaFuel can be produced in any desired amount, whether
small or
large, and anywhere, whether on land or at sea;

        12) AquaFuel is cheaper, simpler and more practical to produce
than
other fuels, such as gasoline, methane or pure Hydrogen;

        13) The AquaFuel technology can be used for recycling waste,
such as
sewer or rubber tires;

        14) The AquaFuel technology can also be used to cleanse the air
from
pollutants;

        15) AquaFuel uses much less Oxygen than other forms of fuels
because of
its high internal content of Oxygen in various compounds, thus providing
a significant contribution to the reduction of the largest environmental
problem of today: the apparent depletion of over 10\% of Oxygen already
occurred so far in our atmosphere, and its exponential increase due to
the excessively large number of automobiles and airplanes all using
Oxygen;

        16) AquaFuel has a light, yet very characteristic and peculiar
odor,
which is an intrinsic characteristic not due to impurities or additives.
This odor has considerable practical and safety values because other
gases which are naturally odorless require the addition of odor via
chemicals for safety purposes.

\section{ Available Measurements}

        Some of the most impressive aspects of AquaFuel are the
simplicity,
safety and rapidity of its production. As an indication, preliminary
measures under verification indicate that the use of 36 V in the arc
permits the production in one minute of 30 liters of AquaFuel at
ordinary pressure. The same production increases exponentially with
voltage, thus permitting the rapid production of AquaFuel anywhere
needed.

     Engine tests of AquaFuel were performed on a small IC engine at the
Briggs \& Stratton Test Center by comparing the new fuel to gasoline. It
was established that AquaFuel requires only five parts of air to one
part of fuel due to its large content of Oxygen bonded in various forms.
By comparison, other gases such as methane require from ten to fifteen
parts of air to one part of gas.

        The results of the Briggs \& Stratton test are as follow:

\vspace{0.3cm}
\begin{tabular}{llll}
  &      Gasoline  &     AquaFuel & \\
       RPM        &      3060     &             3060 & \\
       Torque     &     3.45      &             3.20 & \\
        HP        &             2.05     &              1.86    &  \\
        Oil Temp. &     227\ F         &        165\ F    & \\
        Exhaust    &    751\ F          &       637\ F  & \\
        Hydrocarbons  &         2436 ppm &      185 ppm & (parts per
million) \\
        CO\%    &       4.343     &             .039 & \\
        CO2\%   &       12.086     &            14.695 & \\
        Oxygen \%  &    0.544     &             7.100 & \\
        Hydrocarbon &   13.367   &              0.001 g/hr & (grams per
hour) \\
       Nitrogen Ox. &   5.921     &             0.002 g/hr & \\
        CO       &      421.141     &           0.002 g/hr &
\end{tabular}

\vspace{0.3cm}
The g/hr figures for CO are most impressive. The engine running
on AquaFuel would have to operate for over 210,000 hours to equal the
amount of CO produced in 24 hours while being fueled by gasoline.

         Equally impressive is the presence of 7.1\% of
Oxygen in the exhaust
which is evidently important to maintain the Oxygen balance in our
atmosphere. This is a clear indication that AquaFuel is not a lien
mixture, but a mix with excess power.

         A preliminary analysis of the chemical composition of AquaFuel
was
conducted by NASA  with the following results:

\vspace{0.3cm}
\begin{tabular}{ll}
                Hydrogen    &           46.483 \\
                Carbon Dioxide  &       9.329  \\
                Ethylene        &               0.049 \\
                Ethane          &               0.005  \\
               Acetylene        &       0.616   \\
                Oxygen          &               1.164  \\
                Nitrogen        &               3.818  \\
                Methane         &       0.181   \\
                Carbon Monoxide  &      38.370  \\
                Total             &             100.015
\end{tabular}

\vspace{0.3cm}
The above analysis does not conform to the performance of AquaFuel and
other experimental evidence.  First, the total BTU content of the gas is
less than 300 according to the above data, while the Briggs test shows
almost 90\% of the torque and horsepower on AquaFuel as compared to that
of the same mass of gasoline.  Also, the presence of free Hydrogen in
the combustion chamber in the amount indicated by NASA would produce
Nitrogen Oxide in an amount much greater than that measured by Briggs \&
Stratton.

        Moreover, comparative permeability tests conducted with AquaFuel
and
other gases have established a behavior which disproves the presence in
AquaFuel of 46\% Hydrogen, as claimed by NASA.

       The test was performed by filling up similar balloons with
Hydrogen, CO, Acetylene, Helium and AquaFuel.  The time needed to fully
deflate the balloons back to their original static state was measured as
follows:

\vspace{0.3cm}
\begin{tabular}{ll}
                Hydrogen  &             Two hours. \\
                CO         &            Eight Hours. \\
        Acetylene    &          Two days.  \\
                Helium      &           One week. \\
                AquaFuel     &          Three to six months. \\
\end{tabular}

\vspace{0.3cm}
As the gases leaked out of the AquaFuel balloon, the diameter
was reduced by about 10 \%.  The other gases still in the balloon were
flammable months into the test.  The AquaFuel balloon-test was conducted
three times.

        In view of the above and other evidence, the chemical analysis
by NASA
cannot be considered conclusive, e.g., because of the destruction of the
original structures during testing.

        Other tests on the chemical structure of AquaFuel have also been
done,
but they have been all inconclusive. As an example, tests indicating
high Nitrogen content are in contradiction with the evidence that much
Nitrogen in the mix would render the fuel a cold burner.

        The ratio between the input and output energy is basically
unsettled at
this writing. In fact, all measures of input and output energies are
under way thus being unavailable at this writing in a final form.

        Nevertheless, the presence of additional sources of energy, via
stimulated nuclear transmutations cannot be excluded at this writing
because of a number of direct or indirect indications, such as that on
the {\it explosion of water mist} beasured since the early part of this
century. These issues require specific experimental measures (see Sect.
5) and theoretical studies (see Sect. 6).

        All these inconclusive aspects, which have been studied until
now by
conventional analytic means, indicate that AquaFuel has new
characteristics beyond existing knowledge. For instance, the compound
with atomic weight 14 which is claimed to be Nitrogen could in reality
be a new composite molecule. A similar situation could occur for other
data.

\section{ Ongoing Experimental Measures}

The most important experimental measures on AquaFuel of current primary
interest to TTL are listed below. Suggestions and/or participations by
interested colleagues would be appreciated.

        1) Measure the energy content of AquaFuel per unit volume in BTU
or
other units. It should be indicated that a number of measures of BTU via
conventional means have failed to provide any scientific answer for
various reasons. As an example, readings of BTU compared to methane were
inconclusive because the former burns with about half of the air
(Oxygen) requirement of the latter, thus voiding the scientific value of
any measure without due thermodynamical consideration of the different
air intakes. Similar unsettled results occurred with other measures.
Innovative means for the needed measures are therefore under study.

        2) Measure the individual isotopes in AquaFuel originating from
distilled water. These measures are requested for AquaFuel produced from
distilled water and 99\% pure graphite so that the initial ingredients
are known. The measures should then identify both the atomic number $A$
(total numbers of protons and neutrons) as well as the nuclear charge
$Z$
(total number of protons) for each constituent of AquaFuel. Note that
the identification of all atomic numbers $A$ alone is grossly
insufficient
for an in depth scientific understanding of AquaFuel because of possible
unsettled alternatives on $Z$ for some of the expected $A$, as it will
be
explained in the next section. Measures conducted until now via gas
chromatography and other conventional means have been inconclusive for
various reasons. First, because it was not clearly resolved whether
certain values of $A$  refer to pure elements, or to clusters of them
and,
second, because of the inability to jointly identify the values of $Z$.

        3) Measure the chemical composition of AquaFuel originating from
distilled water. Upon identification of the individual isotopes, or
jointly with the same, the next major measures requested by TTL are the
identification of the chemicals that actually compose AquaFuel. Again,
various tests done until now have resulted to be without final
scientific value, e.g., because they were on based on personal beliefs
by the experimenter disproved by evidence. As an example, the belief
that AquaFuel is primarily composed by Hydrogen has been disproved by
comparative tests on the permeability of Hydrogen and AquaFuel, and
similar unsettled occurrences resulted in other measures.

        4) Measure the chemical composition of AquaFuel
originating from water
inclusive of waste to be recycled. This additional knowledge is
essential for the industrial application of AquaFuel as a means of
recycling contaminated liquids. Various specific cases of recycling are
under consideration for measures.

        5) Identify the chemical structure of the exhaust following
combustion.
This additional knowledge is essential to understand the thermodynamics
underlying the use of AquaFuel.

        6) Identify the physical characteristics of AquaFuel,
such as specific \newline
density. A knowledge of these characteristics is evidently a necessary
pre-requisite for indepth scientific studies on AquaFuel.

        7) Identify compressibility of AquaFuel to the liquid state. A
knowledge of the physical characteristics of temperature, pressure,
etc., for the compressibility of AquaFuel to the liquid state is needed,
as an evident pre-requisite for its storage in tanks.

        8) Identify the structure of the electric discharge
in AquaFuel. It is
generally believed that electric discharges are composed of an intense
flow of individual electrons. As explained in the next section, there
are reasons to suspect that this is not necessarily the case for
AquaFuel because of the alternative possibility already established in
superconductivity that the electric discharge could be composed by an
intense flow of electron pairs. The possible detection of the latter is
evidently important, not only for an indepth scientific study of
AquaFuel but, also, as a foundation of a new technology.

        9) Optimize AquaFuel. On the achievement of at least the
most important
measures identified above, an important objective of TTL is the
maximization of the characteristics of AquaFuel depending on the
application at hand, such as the optimization of the energy/BTU content
for combustion, the minimization or elimination of harmful by-products
in recycling applications, etc.

\section{ Applications of AquaFuel }

The applications of AquaFuel as a combustible gas which have been
identified until now are truly numerous and can be divided into three
classes:

\vspace{0.2cm}
        APPLICATIONS OF CLASS I: FUEL. In this class we note:

        I.1)    Motor fuel because of the remarkable reduction of
pollutants in    the exhaust, high energy content, better
safety, and other aspects        indicated earlier;

        I.2)    Heating fuel for homes and industries, for the same
reasons;

        I.3)    Cooking fuel, because clearly preferable over methane
and
other    gases;

        I.4)    Industrial fuel for a variety of uses, such as for the
furnaces of             the steel industry and others;

        I.5)    Emergency fuel, for instance, for the continuation of
service in a            broken pipeline of natural gas;

        And others.

\vspace{0.2cm}
        APPLICATIONS OF CLASS II: SERVICE. In this class we note:

        II.1) Production of electricity for various industrial and
consumer         uses;

        II.2) Recycling of liquid waste such as sewage;

        II.3) Recycling of solid waste such as rubber tires;

        II.4) Environmental clean--ups;

        II.5) Desalination;

and other uses.

\vspace{0.2cm}
        APPLICATIONS OF CLASS III: PROCESSING. In this class we
note:

        III.1) Separation of water;

        III.2) Production of new chemicals;

        III.3) Production of gases;

and other possibilities currently under study.

        In summary, AquaFuel is a new combustible gas with undeniably
less
pollutant when compared to other fuels of current uses, it is very
simple to be rapidly produced in large volumes, and it exhibits features
which are clearly beyond pre--existing theories.  As such, AquaFuel not
only has many valuable characteristics and applications available now,
but also constitutes a sound platform for systematic research leading to
new technologies.

\section{
The New Methods Needed For Scientific
Studies Of New Energies}

Quantitative scientific studies of new forms of energy in general, and
of AquaFuel in particular,  require the consideration of all possible or
otherwise expected profiles, including studies within the context of
particle physics, nuclear physics, chemistry, thermodynamics,
superconductivity,  and other disciplines.

        Rather than being a drawback,
this occurrence is important because the
conduction of in depth, mathematical, theoretical and experimental
studies on new forms of energy is expected to permit the understanding
of their origin, which is an evident pre--requisite for the improvement
of their efficiency in a systematic scientific way, rather than via the
current empirical process of endless {\it trials and errors}.

        It should be indicated from the outset that the use of
conventional
theories, such as Quantum Mechanics and Chemistry, has not permitted a
scientific resolution of the various issues one way or the other, and
has only caused unresolved controversies (see, e.g., the related
articles on AquaFuel and other topics in Refs. [2]).

        It is today well established that the sole use of these theories
has no
longer a final scientific value for new energies. In fact, these
energies, to be truly {\it new}, by definition and conception, have to
be
{\it beyond} said conventional theories. This is due to the fact that
quantum mechanics and chemistry have well defined limitations, while the
processes underlying  new energies clearly go beyond these limitations.

        Unfortunately, the limitations of quantum mechanics and
chemistry are
{\it the best kept secrets of the trade} in the sense that they are
known to
true experts in the field, but never disclosed during Ph.D. classes or
in orthodox technical papers. In fact, the very admission of these
limitations is a de facto admission of the need for more adequate
theories, and this is notoriously against established interests.

        A systematic study of these limitations has been recently made
available by this author in monographs [4]. As a nontechnical outline,
we here mention that quantum mechanics and chemistry have a structure
which is linear (i.e., only dependent on the first power of the
wavefunctions), local  (i.e., solely defined for a finite set of
isolated points) and potential  (i.e., only admitting
action--at--a--distance interactions derivable rom a potential energy).

        As a consequence, quantum mechanics and chemistry can only
represent
systems as being composed of point--like particles with
action--at--a--distance interactions. As an example, the valence
electrons
of the molecules used in new forms of energy (e.g., the water used by
AquaFuel) are represented in contemporary quantum chemistry as
dimensionless points.

        Though of unequivocal scientific value, this is evidently only
an
approximation of nature because, in the physical reality, all particles,
including the valence electrons, have an extended wavepacket of the
order of $10^{-13}$ cm.

        When the particles are at the (relatively) large mutual
distances of an
atomic structure ($10^{-8}$ cm -- which is 10,000 times bigger than the
size
of the wavepackets of the electrons), their extended character can be
effectively ignored and the resulting theories are exact.

        The best known example of the exact validity of quantum
mechanics is
the atomic structure. In fact, quantum mechanics has represented in a
numerically exact way the totality of experimental data on the structure
of the Hydrogen atom.

        The best known example of the limitations of quantum mechanics
and
chemistry is the study of the two Hydrogen atoms composing the Hydrogen
molecule because:

        1) Quantum chemistry cannot represent the strongly attractive
force
between the two atoms since they are neutral, thus implying null average
Coulomb forces among the atoms;

        2) Quantum chemistry has been unable to represent throughout
this
century 100\% of the binding energy of the Hydrogen molecule;

        3) Quantum chemistry cannot explain why the
Hydrogen and water
\newline molecules have only two atoms, since it
has been proved to admit an arbitrary number of atoms;

        4) Quantum chemistry cannot permit meaningful
 thermodynamical studies
on reactions based on the Hydrogen and water molecules
because the 2\% of
binding energy which is not accounted for is
deceptively small, since it
corresponds to about 950 Kilocal/mole while an
ordinary reaction
requires about 20 Kilocal/mole;

        5) Thermodynamics predicts a behavior of the Hydrogen and water
\newline molecules under electric and magnetic fields
dramatically against experimental evidence (e.g., the
capability for the molecule to be magnetized like an ordinary
ferromagnet); and other shortcomings.

        The origin of the above shortcomings is in the structure of
molecules
themselves, where we have valence electrons which can interact at mutual
distances of the order of the size of their wavepackets, which is a
distance 108 times smaller than the atomic distances. Under the latter
conditions the extended size of the wavepacket of the particle is no
longer ignorable, thus implying a clear limitation of quantum mechanics
and chemistry.

        In fact, the deep overlapping of the wavepacket of particles
implies
new interactions which are technically called nonlinear  (in the sense
of depending on powers of the wavefunctions bigger than one), nonlocal
(in the sense that they are extended over the volume of
wave-overlappings which cannot be evidently reduced to finite number of
isolated points), as well as nonpotential (in the sense of being of
contact / zero--range type for which the notion of
action--at--a--distance
potential has no mathematical or physical meaning of any type).

        As such, the representation of the deep overlapping of the
wavepackets
of valence electrons as well as of other nuclear events is beyond any
credible  hope of quantitative treatment via quantum mechanics or
chemistry on conceptual, mathematical and physical grounds.

        To put it explicitly, any treatment of truly new  forms of
energy via
quantum mechanics and chemistry is not scientific if considered as of
{\it final} character because these disciplines approximate everything
as
{\it points}, thus being unable to represent in any credible way the
main
mechanisms which are at the very foundation of the new  energies.

        Comprehensive studies on the structural generalization of
quantum
mechanics for quantitative treatments of nonlinear, nonlocal and
nonpotential interactions at short distances were initiated by R. M.
Santilli at Harvard University back in 1978 under support from the U.S.
Department of Energy [3]. The studies were then continued by an
increasing number of scientists, and have resulted in over 10,000 pages
of published research, including about one thousand papers published in
numerous mathematics and physics journals all over the world, some 20
advanced (post Ph. D.) monographs and about 40 volumes of proceedings of
international conferences held in the USA and Europe (see monographs
[4]). As a result of these collegial efforts  Hadronic Mechanics has
today reached operational maturity (see the recent 104 page long memoir
[5]).

        The new mechanics was called by Santilli Hadronic Mechanics  to
characterize the description of strongly interacting particles, such as
the nuclear constituents, collectively called {\it hadrons}, under
conventional long-range potential, as well as short--range, nonlinear,
nonlocal and nonpotential interactions.  It should be indicated that
Hadronic Mechanics is the only  mechanics available today for the
consistent treatment of the new nonlocal and nonpotential interactions
due to deep wave-overlappings. Other generalized theories have
well--known physical inconsistencies, such as the inability to have
unique and invariant numerical predictions.

        Note that Hadronic Mechanics coincides with quantum mechanics at
large
distances. Moreover, Hadronic Mechanics preserves all conventional
quantum laws, such as Pauli's exclusion principle, Heisenberg's
uncertainty law, etc., and only realize them in a more general way.

        The only differences occur at distances of the order of
$10^{-13}$ cm.
As a result, Hadronic Mechanics has achieved Einstein's historical
{\it completion} of quantum mechanics with new, generally small
contributions at short range.

        Hadronic Mechanics has permitted the construction of a
broadening of
quantum chemistry called Hadronic Chemistry [12] (also called
{\it isochemistry} for certain technical treasons). Again, quantum and
hadronic chemistry coincide everywhere except at short distances where
the new chemistry includes the nonlinear, nonlocal and nonpotential
effects which are absent in the older theory.

        A main feature of the Hadronic Mechanics and Chemistry which is
important for all new  forms of energy, including AquaFuel, is that:

When two particles couple themselves in a singlet  at short distances
(one with spin up and one with spin down) the new  nonlinear, nonlocal
and nonpotential forces due to deep wave--overlappings are so attractive
to overcome possible repulsive Coulomb forces, and permit new bound
states at short distances which simply cannot be conceived - let alone
treated -- by quantum mechanics and chemistry.

                The sphere with radius 1 fm $= 10^{-13}$ cm in
which Hadronic Mechanics and
Chemistry have the novel effects is called hadronic horizon. In the
outside of the hadronic horizon conventional quantum mechanics and
chemistry hold, while in its interior the broader hadronic mechanics and
chemistry hold. Conventional physical laws (such as Pauli's exclusion
principle, total conservation laws, etc.) hold everywhere including in
the interior of the hadronic horizon.

        If particles have opposite charges, they naturally penetrate
such a
horizon (when in singlet coupling and under a number of other
conditions, such as the conservation of energy and other total
quantities) by therefore activating the new effects rather
spontaneously. This is the case of electrons and positrons which
naturally attract each other and can apparently form a new bound state
at short distance (in addition to the known electronium at large
distances) commonly known as the neutral $\ast$--meson [3].

        On the contrary, if particles have the same charges, they can
penetrate
the hadronic horizon only under the assistance of an outside  trigger,
namely external conventional interactions which favor the particles to
move toward each other against their repulsive Coulomb force. This is
the case of the coupling of the valence electrons in molecular bonds in
which the {\it trigger} is given by the nuclei of the molecules which,
being
positively charged, attract the the valence electrons to such short
distances to permit their bond [12].

        Systematic studies of the physical laws of new forms of energy
as
predicted by hadronic mechanics and chemistry are under way [11]. We
outline below some of the advances permitted by the new theories which
are playing a basic role in these studies. In fact, hadronic mechanics
and chemistry have permitted:

                1) The first theoretical representation of
the bond of two identical
electrons in the Cooper pair of superconductivity [6]:

\begin{equation}
   e  +  e   +   \mbox{trigger}    \to     \mbox{CP}
\qquad \mbox{(Cooper Pair at $10^{-8}$ cm)}\ ,
\end{equation}
which is in excellent agreement with experimental data in
superconductivity, where the {\it trigger} is given by the Cuprate ions
(see
Refs. [6] for technical details, including the clear emergence of
nonlinear, nonlocal and nonpotential effects as a necessary condition
for the attraction).

        The scientific scenario can be here identified in a way so clear
to
un-mask possible nonscientific postures in favor of old theories for
personal gains.

        On one side, physical reality establishes that electrons can
bond
themselves, not only in the Cooper pair (where the bond is so strong
that the coupled electrons have been detected to tunnel together through
a potential barrier), but also in the Helium (where the two orbital
electrons often travel bonded together rather than isolated), as well as
in ball lightning (in which electrons bond together in very large
numbers as per incontrovertible evidence, this time visible by the naked
eye).

        On the other side, quantum mechanics and chemistry simply cannot
represent this physical reality at the level of individual electron
pairs because of the Coulomb repulsion (they do provide a representation
but only at the statistical  level of an ensemble of pairs  which is
absolutely not the issue here, since we are referring to specific
differential equations of structure clearly exhibiting an attraction for
one single and individual pair of electrons  [6]).

        Ergo, quantum mechanics and chemistry have limitations which
cannot be
denied as a condition for scientific credibility. The moment these
limitations are admitted, and only then, the door is open to truly basic
advances.

        Hadronic Mechanics and Chemistry do indeed permit a
quantitative-numerical representation of the {\it attractive} force
between
the {\it identical} electrons in the Cooper pair, the Helium, the ball
lightning and other events in excellent agreement with experimental data
[6].

        By no means these results are expected to be unique
because a beauty of
real science is its polyhydric character. Studies with alternative
approaches are not only welcome but actually solicited, provided  that
they achieve similar results in an invariant form.

        2) A new model of the Hydrogen molecule with the first
explicitly
attractive force between neutral atoms due to the pairing of the valence
electrons into a hadronic bound state at short distances called
{\it isoelectronium} [12],

\begin{equation}
          e  +   e   +   \mbox{trigger}   \to
\mbox{IE}   \qquad \mbox{ (Isoelectronium at $10^{-11}$ cm})\ ,
\end{equation}
where the trigger is in this case given by the positive nuclei of the
Hydrogen molecule.

        Hadronic Mechanics and chemistry have resolved all the
shortcomings of
quantum chemistry indicated earlier, by reaching for the first time: an
explicit attractive force among the neutral atoms of the molecular bond
due to the short range bonding of the valence electrons into a single
state; a representation of 100\% of the experimental data; a
quantitative
explanation of why only two Hydrogen atoms are admitted in the molecule;
and other advantages over  quantum chemistry.

        3) A new model of the water molecule which resolves at least
some of
the existing problematic aspects [12]. The reader should be aware that
despite the conduction of studies since the initiation of science, the
structure of water remains vastly known at this writing.

        In fact quantum chemistry: a) does not admit a significant
attractive
force among the neutral atoms composing the molecule (the esoteric
forces currently believed to yield the bond, such as the exchange or van
der Waal forces, have been rigorously proved to yield very small, thus
insufficient attractions); b) has been unable to represent 100\% of the
experimental data on the binding energy, electric and magnetic moments
and other aspects; c) has been proved to admit admit an arbitrary number
of Hydrogen atoms in flagrant disagreement with evidence; d) cannot
permit meaningful thermodynamical studies due to the excessively large
error in thermodynamical units recalled earlier; e) predicts a behavior
of water under magnetic and electric fields which is in dramatic
disagreement with experimental evidence (e.g., it has been proved via
the use of quantum electrodynamics that water under a magnetic field
should acquire a net North-South polarity as for an ordinary
ferriomagnet, in gross disagreement with experimental evidence); and
possesses other, generally unknown insufficiencies.

        It is evident that, under all these insufficiencies, no study of
new
forms of energy using water, such as AquaFuel, SkyGas, water explosions,
etc., can be considered as {\it final}.

        It is evident that the true scientific evidence is the
limitation of
quantum chemistry for truly scientific studies of these new forms of
energy.

        Again, once these limitations are admitted, the door is open to
basic
advances. Hadronic Chemistry resolves the above insufficiencies by
permitting a deeper (although never {\it final}) understanding of water
[12]. Other approaches are welcome and encouraged.

        4) The first theoretical representation of the synthesis of the
neutron
as occurring in stars at their formation, namely, from protons and
electrons only (because at their formation stars are solely made up of
Hydrogen) according to the reaction [7]
\begin{equation}
p  +  e    \to     n  + \mbox{neutrino} \ .
\end{equation}

The above reaction verifies all known physical laws. Nevertheless, its
rate (cross section) according to quantum mechanics is very small and,
therefore, the reaction is believed not to have practical value.
Hadronic mechanics recovers the above low rate for all energies, except
at a specified threshold energy in which the rate of reaction (3) has a
large peak.

        Note that quantum mechanics cannot represent the neutron as a
bound
state of a proton and an electron only because of a host of
inconsistencies all resolved by the Hadronic Mechanics and Technologies.

        We should finally note that, though preliminary and in need of
independent re-runs, an experimental verification of synthesis (3) was
successfully conducted in Brasil [8].

        5) The first theoretical representation of the stimulated decay
of the
neutron [9] according to the reaction
\begin{equation}
 \gamma +  n     \to    p
+  e   +  \mbox{anti--neutrino}\ .
\end{equation}

Again, the above reaction verifies all known physical laws. Quantum
mechanics predicts that its rate is very small for all possible energies
of the photon, thus having no practical value. Hadronic Mechanics and
Technologies recover this feature, but predict the existence of a large
peak in the rate of reaction (4) at a specific resonating frequency,
thus permitting the stimulated decay of the neutron in a form suitable
for industrial applications.

        It should be noted that the proton is stable and, as such,
cannot be
stimulated to decay according to current knowledge. On the contrary, the
neutron is naturally unstable. The identification of means for its
stimulated decay is, therefore, only a matter of time.

        We should finally note that, though preliminary and in need of
independent re-runs, an independent experimental verification of
synthesis (3) was successfully conducted in Greece [10].

        6) Stimulated nuclear transmutations due to an electron capture
by
certain nuclei [11]. Nature establishes the existence of the spontaneous
electron capture  (EC), a process in which a given nucleus $(A,\ Z)$
with
total number of protons and neutrons $A$ and total number of protons $Z$
(the nuclear charge) absorbs one electron from the environment or from
the peripheral atomic cloud. During this process the atomic number $A$
evidently remains the same, but $Z$ decreases  by one unit because we
have
the synthesis of one neutron from a proton and an electron as per
reaction (3), thus resulting in the nuclear reaction
\begin{equation}
 (A,\ Z)  +  e  +
\mbox{trigger}     \to    (A,\ Z-1) \ ,
\end{equation}
where the trigger  represents all the conditions needed to allow the
reactions, such as the energy missing in order to verify the basic
principle of conservation of the energy.

        Since the above reaction occurs spontaneously in nature (under
the
right circumstances), Hadronic Mechanics and Technologies predict that
it can be stimulated under suitable conditions.

        Note that transmutation (5) is predicted at a sub-nuclear level,
that
is, it should be studied as a process in the interior of one-nuclear
constituent, because the same transmutation is not possible when studied
at the nuclear level, that is, at the level of a nucleus as a collection
of protons and neutrons.

        7) Stimulated nuclear transmutations due to neutron decay [11].
If
reaction (4) occurs in nature for an isolated neutron, it must evidently
occur also when the same neutron is a member of a nuclear structure,
resulting in the stimulated nuclear decay
\begin{equation}
  (A,\ Z)  +  \gamma +  \mbox{trigger}
\to      (A,\  Z+1)\ .
\end{equation}

        Note, again, that the above stimulated nuclear transmutation can
only
occur at one, single, fixed resonating frequency, and can only occur at
the sub-nuclear level, rather than at the level of a nucleus as a
collection of protons and neutrons.

        8) Stimulated nuclear transmutations due to proton capture [11].
Physical evidence establishes that two identical electrons can bond
themselves into one single state, the Cooper pair in superconductivity,
Eq. (1). A similar occurrence exists in the bond of the valence
electrons in molecular structures, Eq. (2). It is evident that this
physical reality is independent of the charge and the mass. As a result,
the experimental evidence on the Cooper pair is de facto evidence on the
expected existence of a similar occurrence for protons.

        Hadronic Mechanics and Technologies therefore predict that,
evidently
under certain conditions, one proton can indeed bond itself to a nucleus
despite their repulsive Coulomb forces, according to the reaction
\begin{equation}
    (A,\ Z)  +  p  +  \mbox{trigger}     \to  (A+1,\ Z+1)\ .
\end{equation}

        We merely have an occurrence equivalent to the experimentally
established creation of ball lightning, which is constituted by
individual electrons bonding themselves to a large number of identical
electrons. According to quantum mechanics, the creation and growth of
such ball lightning is not allowed because of the Coulomb repulsion. Yet
the phenomenon exists in the physical reality. Therefore, to do science,
we have to  accept physical evidence and modify the theory to reach its
quantitative interpretation, rather than trying to adapt physical
evidence to pre-existing theories. Hadronic Mechanics and technologies
do precisely that, by surpassing pre-existing theories to accommodate
experimental facts.

        The case on one proton bonding itself to a nucleus under the
necessary
conditions is fully equivalent in all respects to the accretion of ball
lightning. The sole difference is that the proton is heavier than the
electrons and the charges of reaction (7) are positive, rather than
negative. The important point is that the underlying physical laws are
exactly the same.

        9) New thermochemical reactions [6]. It is evident that the
availability of a new model of molecular bonds in general, and of the
structure of water in particular, imply new chemical features and
reactions which, as it is the case for stimulated nuclear transmutations
(5), (6) and (7), simply cannot be predicted, let alone treated via
quantum chemistry. One of these predictions which has already been
verified experimentally is a novel behavior of the water and new
chemical reactions under electric and magnetic fields with consequential
release of new forms of energy.

        It is evident that all  the above new advances 1) -- 9) are
potentially
applicable to new forms of energy, including AquaFuel, and they will
likely identify new technologies.

        The theoretical research under way therefore contemplates the
systematic application of the above new advances to various aspects of
new energies including:

        A) Particle physics, such as the possibility that electric
discharges
are made up of isoelectronium pairs, rather than individual electrons as
currently believed until now;

        B) Nuclear physics, such as the expected nuclear transmutation
due to
stimulated electron capture, decay of the neutron from the intense light
in the discharge, accretion of protons in the completely ionized plasma,
etc.;

C) Chemistry, such as possible new chemical bonds and new
chemical reactions with related release of new energies that are
inconceivable for quantum chemistry;

        D) Thermodynamics, such as the new energy calculations which are
not
possible at the moment due to the excessive error indicated earlier;

E)
Superconductivity, because an electric discharge is the ideal limit case
of superconductivity;

and other aspects.

        We hope to report in Infinite Energy  the most salient advances
on the
above research for the benefit of all new forms of energies.

        The content of this paper first appeared in Hadronic Journal
Supplement, Vol. 13, pages 1--22, 1998.

\section*{ Acknowledgments.}
The author would like to thank William H. \newline Richardson, jr.,
Jerry
Kammerer, Mark Clancy, Ken Lindfors and all other members of Toups
Technology Licensing, Inc.,  as well as Rob Jaeger, Peter Gluck, Dieter
and Yvette Schuch, and various colleagues, for invaluable critical
comments on earlier versions of this paper.


\begin{thebibliography}{00}

\bibitem{r1.}      W. H. Richardson, jr., US Pat 5,435,274
"Electrical power generation
         without harmful emissions", July 25, 1995;    US 5,692,459
"Pollution      free vehicle operation", Dec 2, (1992); patents
pending.

\bibitem{r2.}      Infinite Energy no. 9, 1996; no. 10, 1996;
no. 11, 1996; no. 13 1997;   and no. 14, 1997; International
Journal of Hydrogen Energy  no. 21,     Feb 1996 p 142 in H.
Roger Hind's column "News and Views"

\bibitem{r3.}      R. M. Santilli, in Hadronic J. no. 1 and 2,
(1978).

\bibitem{4.}      R. M. Santilli, Elements of Hadronic
Mechanics, Vols. I and II,  Ukraine Academy of Sciences, Kiev,
second edition, (1995).

\bibitem{5.}      R. M. Santilli, Found. Phys. Vol. 27, 625,
(1997).

\bibitem{6.}    A. O. E. Animalu, Hadronic J. no. 17, p. 379
(1995). A. O. E.  Animalu and R. M. Santilli, Intern. J. Quantum
Chemistry 29,   175  (1995).

\bibitem{7.}    R. M. Santilli, JINR Comm. no. E4-93-352, Dubna,
Russia; and     Chinese J. Syst. Eng. \& Electr. (patent pending).

\bibitem{8.}    C. Borghi et al. (Russian) J. Nuclear Physics Vol.
56, 147, (1993).

\bibitem{9.}    R. M. Santilli, in the Proceedings of
International Symposium on  Large Scale Collective Motion of
Atomic Nuclei, G. Giardina et al.,  Editors, World Scientific,
Singapore (1997) (patent pending).

\bibitem{10.}   N. F. Tsagas et al., Hadronic J. Vol. 19, 87
(1996).

\bibitem{11.}   R. M. Santilli, "Physical laws of the emerging new
energies  as   predicted by hadronic mechanics and chemistry",
I, II and III IBR preprints RMS-003, 004 and 005 (1998)(patent
pending).

\bibitem{12.}   R. M. Santilli and D. Shillady, "Ab initio
isochemistry", I and  II,       IBR preprints 1998 submitted for
publication. R. M. Santilli and D. D.   Shillady, Hadronic J.
vol. 21 (1998), in press.
\end{thebibliography}
\end{document}